\documentclass[aps,pra,reprint,superscriptaddress,floatfix]{revtex4-2}

\usepackage{graphicx}
\usepackage{float}
\usepackage{subcaption}
\usepackage{amsmath}
\usepackage{amssymb}
\usepackage{hyperref}
\usepackage{braket}
\usepackage{xcolor}
\usepackage{placeins}
\usepackage{algorithm}
\usepackage{algpseudocode}

\newcommand{\G}{\mathcal{G}}

\renewcommand{\O}[1]{\mathcal{O}\left(#1\right)}

\renewcommand{\bf}[1]{\mathbf{#1}}

\begin{document}

\title{Approximation and composition of functions in quantized tensor trains via orthogonal polynomial expansions}

\author{Juan José Rodríguez-Aldavero}
\email{jj.rodriguez@iff.csic.es}
\affiliation{Institute of Fundamental Physics IFF-CSIC, Calle Serrano 113b, Madrid 28006, Spain}
\author{Paula García-Molina}
\affiliation{Institute of Fundamental Physics IFF-CSIC, Calle Serrano 113b, Madrid 28006, Spain}
\author{Luca Tagliacozzo}
\affiliation{Institute of Fundamental Physics IFF-CSIC, Calle Serrano 113b, Madrid 28006, Spain}
\author{Juan José García-Ripoll}
\affiliation{Institute of Fundamental Physics IFF-CSIC, Calle Serrano 113b, Madrid 28006, Spain}

\date{19 August 2026}

\begin{abstract}
    This work explores the representation of univariate and multivariate functions as quantized tensor trains (QTT). It develops a constructive algorithm that employs expansions in orthogonal polynomial bases and Clenshaw evaluations to represent analytic and highly differentiable functions as QTT approximants. This approach provides a general framework for function composition in QTT form that generalizes efficiently to multidimensional settings. Without loss of generality, the presentation focuses on the Chebyshev basis, where the method demonstrates rapid convergence for highly differentiable functions, in agreement with classical theoretical predictions, and remains numerically stable at high polynomial orders. The performance of the algorithm is compared with that of tensor cross-interpolation (TCI) and multiscale interpolative constructions, demonstrating competitive performance and favorable scaling behavior in certain scenarios.
\end{abstract}

\maketitle

\section{Introduction}
\label{sec:1_introduction}

Low-rank tensor decompositions~\cite{kolda2009}, also known as tensor networks~\cite{orus_2014}, provide a powerful platform for representing and manipulating high-dimensional objects arising in numerical analysis. Among these, the tensor-train (TT) decomposition~\cite{oseledets_2011} and its quantized variant, the quantized tensor-train (QTT)~\cite{khoromskij_2011}, enable the compressed representation of functions and operators defined on large tensor-product grids by exploiting hierarchical low-rank structure. When such a structure is present, QTT representations can achieve exponential reductions in storage and computational complexity compared to explicit grid-based discretizations.

Tensor-train representations were originally introduced in the context of strongly correlated quantum systems under the name of matrix product states (MPS)~\cite{affleck.etal_1987, fannes.etal_1989}, and gave rise to highly successful algorithms such as the density matrix renormalization group (DMRG)~\cite{white_1992, ostlund.rommer_1995,dukelsky.etal_1998}. Since then, closely related tensor-network techniques have been adopted in numerical analysis under the umbrella of \emph{quantum-inspired} numerical methods~\cite{garcia-ripoll_2021}, with applications including Fourier analysis~\cite{dolgov.etal_2012a, garcia-ripoll_2021, chen.etal_2023, garcia-molina.etal_2022} and solving high-dimensional differential equations~\cite{dolgov.etal_2012, ye.loureiro_2022, gourianov.etal_2022, garcia-molina.etal_2022}. Furthermore, the interpretation of tensor network formalisms as quantum circuits~\cite{schon.etal_2005} makes them an efficient method for discovering quantum algorithms, thus interfacing with quantum computing applications~\cite{melnikov.etal_2023}.

A central challenge in QTT-based numerical methods is the construction of accurate tensorized representations of functions. A direct approach, based on applying the TT-SVD algorithm to the full tensor-product discretization~\cite{oseledets_2011}, scales exponentially with dimension and is therefore limited to low-dimensional scenarios. Existing alternatives can be broadly classified into three categories. The first relies on adaptive sampling strategies that construct tensor approximations directly from black-box function evaluations, most notably algorithmic variants of tensor cross-interpolation (TCI)~\cite{oseledets.tyrtyshnikov_2010,savostyanov.oseledets_2011,dolgov.savostyanov_2020,ritter.etal_2024}. A second category comprises randomized projection or sketching methods, which approximate structured tensors via randomized or structured linear maps~\cite{hur.etal_2023, monturiol.etal_2025}. A third class exploits explicit analytical or algebraic structure in the target function, enabling constructive tensor representations based on polynomial expansions or algebraic mappings~\cite{peng.etal_2023,lindsey_2023,chen.lindsey_2024}.

In this work, we propose a method belonging to the latter category, based on expanding functions in orthogonal polynomial bases and constructing the resulting linear combinations directly in QTT format through a finite-precision algebra of operations~\cite{szeg1939}. The core algorithm combines polynomial interpolation with Clenshaw-type recurrence relations and variational TT compression to obtain low-rank QTT approximations of analytic and highly differentiable functions~\cite{garcia-ripoll_2021,garcia-molina.etal_2022}. Without loss of generality, the presentation and numerical experiments focus on the Chebyshev basis~\cite{mason.handscomb_2003,trefethen_2019, halimeh.etal_2015,soley.etal_2022}, while the framework naturally extends to general families of orthogonal polynomials.

The proposed construction expands a univariate function in an initial argument represented in QTT form, which may itself encode a multivariate function, enabling function composition within the QTT format. This scheme allows for the approximation of functions with compositional algebraic structure. The construction proceeds incrementally in the polynomial degree, with rank truncation applied at each step to control rank growth. In all tested scenarios, the method exhibits rapid convergence for highly differentiable functions, consistent with classical Chebyshev approximation theory, and proves numerically stable even at large interpolation orders. Furthermore, it generalizes efficiently to multivariate settings while maintaining favorable scaling rates.

We benchmark the performance of the algorithm on a collection of univariate and multivariate target functions and compare it with existing QTT approximation techniques, including TCI and multiscale interpolative constructions. While the method generally does not outperform TCI in univariate settings when function evaluations are inexpensive or the target lacks smoothness, it provides a deterministic, structure-preserving alternative grounded in classical polynomial approximation theory. In smooth regimes, it often requires only a small number of univariate samples, several orders of magnitude fewer than purely sampling-based approaches. Moreover, unlike multiscale interpolative constructions that directly approximate the full high-dimensional function, the proposed method scales efficiently to multivariate settings by exploiting compositional structure. This leads to a tunable accuracy-cost tradeoff and competitive performance in high-dimensional examples involving up to 200 QTT cores, corresponding to tensors with $2^{200}$ entries, far beyond the reach of conventional dense representations.

This work is organized as follows. Section~\ref{sec:2_qtt_framework} introduces the QTT representation of functions and the finite-precision QTT algebra used throughout the paper. Section~\ref{sec:3_orthogonal} presents the QTT orthogonal polynomial expansion algorithm and its realization for the Chebyshev basis. Section~\ref{sec:4_results} reports univariate and multivariate numerical benchmarks, including Gaussian-like multivariate functions and quartic models. Finally, Section~\ref{sec:5_discussion} discusses the implications of the results and outlines directions for future work.

\section{Quantized tensor trains}
\label{sec:2_qtt_framework}

\begin{figure}[t]
  \centering
  \includegraphics[width=0.8\columnwidth]{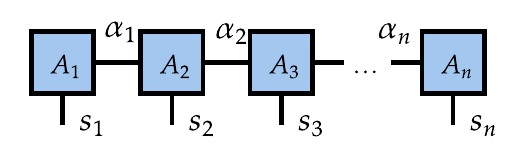}
  \caption{QTT composed of $n$ mutually contracted order-3 tensors $A_k$, with contracted indices $\alpha_k$ and free indices $s_k$.}
  \label{fig:qtt}
\end{figure}

Tensors are central objects in multilinear algebra and numerical analysis. When represented as multi-index arrays $A_{s_1 s_2 \ldots s_n}$ with $s_i \in \{0, \ldots, \ell_i - 1\}$, their storage and manipulation costs grow exponentially with the order $n$, rendering direct numerical treatment intractable in high-dimensional settings.

In the tensor-train (TT) representation, tensors are decomposed into a collection of mutually contracted, lower-order tensors known as TT cores. Denoting regular tensors by capital letters and tensor trains using boldface letters, this decomposition reads
\begin{equation}
  \bf{A}_{s_1 s_2 \ldots s_n} = \sum_{\alpha_1=1}^{r_1}\sum_{\alpha_2=1}^{r_2}\dots \sum_{\alpha_n=1}^{r_n} A_{\alpha_1}^{s_1} A_{\alpha_1 \alpha_2}^{s_2} \dots A_{\alpha_{n-1}}^{s_n},
  \label{eq:tt}
\end{equation}
with the $r_k$ known as the ``TT ranks'' or ``bond dimensions'', in such a way that each of the individual tensor elements is determined by a product of matrices. Following Penrose's graphical notation (see Fig.~\ref{fig:qtt}), this decomposition is depicted as a one-dimensional train of boxes, where each box corresponds to a tensor, the lines that connect them represent the contracted indices $\alpha_k$, and the free indices $s_k$ become freely standing ``legs''.

By construction, the number of elements in a TT grows polynomially with the number of components $\O{n\ell r_{\max}^2}$, where $r_{\max}:=\max r_k$ is the largest TT rank. When these ranks remain bounded independently of $n$, the TT representation becomes exponentially more efficient than the full tensor $\bf{A}_{s_1 s_2 \dots s_n}$. Let us now discuss how these improvements apply to the representation of functions.

\subsection{QTT representation of functions}
\label{sec:2_qtt_framework_functions}

The quantized tensor-train (QTT) formalism interprets a discretized function as a tensor whose indices label the grid points~\cite{khoromskij_2011}. For a one-dimensional continuous variable $x$ defined in an interval $[a, b]$, a set of $n$ indices running over $\ell$ values, $s_i \in \{0, 1, \ldots, \ell-1\}$, can label up to $\ell^n$ points. In particular, when the grid is uniform, a $\ell$-nary representation of the points is given by
\begin{equation}
    x_{s_1 s_2 \ldots s_n} = a + (b-a) \sum_{i=1}^n s_i \ell^{-i}.
\end{equation}
These $\ell^n$ values may be regarded as a tensor with $n$ indices, $s_1$ to $s_n$, each addressing progressively smaller length scales. The tensor encoding of the $x$ coordinate generalizes to arbitrary univariate functions,
\begin{equation}
    f_{s_1 s_2 \ldots s_n} := f(x_{s_1 s_2 \ldots s_n}),
\end{equation}
and extends to $m$-dimensional functions, with the caveat that we now have multiple choices for the ordering of indices~\cite{garcia-ripoll_2021}. In this work, we consider either a sequential (serial) order
\begin{equation}
  f_{s_1^1s_2^1\ldots s_{n_1}^1\ldots s_1^m s_2^m\ldots s_{n_m}^m}=f(x^{1}_{s_1 s_2\ldots s_{n_1}}, \ldots, x^{m}_{s_1 s_2\ldots s_{n_m}}),
  \label{eq:f_sequential}
\end{equation}
or an interleaved order $f_{s_1^1\ldots s_1^ms_2^1\ldots s_2^m\ldots s_{n_1}^1\ldots s_{n_m}^m}$, corresponding to a reordering of the bits in the sequential layout~\eqref{eq:f_sequential}. All these tensors are amenable to a QTT decomposition~\eqref{eq:tt}, with a complexity determined by their TT rank structure. Their compressibility depends on properties such as their algebraic form, smoothness, or displacement structure~\cite{shi.townsend_2021,griebel2022}.

Smooth univariate functions are known to be highly compressible, as they yield QTTs with provably bounded TT ranks~\cite{garcia-ripoll_2021,jobst.etal_2023}. For instance, polynomials of finite degree $d$ admit an exact representation with $r_{\max} = d+1$~\cite{lindsey_2023}. A general recipe for constructing such representations is given in~\cite{oseledets_2013} and, alternatively, in~\ref{sec:A_polynomials} using an inductive argument. Unfortunately, these polynomials rely on the monomial basis and are thus ill-conditioned for function approximation. Other canonical examples include sums of $d$ exponentials, which have TT ranks equal to the number of summands, $r_{\max} = d$, yielding exact sine and cosine representations with $r_{\max}=2$ on uniform grids.

In the multivariate setting, much of the theoretical effort has focused on tree tensor networks (TTN), also known as hierarchical Tucker decompositions. These structures use a hierarchical tensor arrangement---unlike the linear topology of QTT---and have been shown to satisfy theoretical complexity bounds under various regularity assumptions~\cite{griebel2022, ali2023c, ali2024a} and algebraic structure~\cite{bachmayr2021}.

In general, loading a function $f$ in a QTT requires a specialized algorithm. A na\"{\i}ve approach discretizes the function on a tensor $f_{s_1 s_2 \dots s_n}$ and applies the TT-SVD decomposition. This creates a QTT with a quasi-optimal error in Frobenius norm~\cite{oseledets_2011}, but requires time and memory that scale with the volume of the tensor---i.e., exponentially in the number of indices $n$. Fortunately, other algorithms exist with a polylogarithmic cost relative to this volume. As mentioned in the introduction, these methods are divided into those that require some randomized or induced sampling---e.g., tensor cross-interpolation (TCI) methods~\cite{oseledets.tyrtyshnikov_2010,dolgov.savostyanov_2020, savostyanov.oseledets_2011,savostyanov_2014, ritter.etal_2024} and sketching methods~\cite{hur.etal_2023, monturiol.etal_2025}---and those that are based on some algebraic structure---such as rank-revealing interpolative constructions~\cite{lindsey_2023} and arithmetic circuit tensor networks~\cite{peng.etal_2023}.

This work proposes a different approximation technique that combines the properties of interpolative formulas~\cite{lindsey_2023} with the algebraic structures of arithmetic circuit tensor networks~\cite{peng.etal_2023}. It is based on a finite-precision non-linear algebra of QTT operations, supporting linear combinations and Hadamard products, which is used to implement a polynomial expansion scheme. The resulting algorithm constructs QTT approximants for univariate functions $f(x)$, and naturally extends to multivariate functions with a compositional algebraic structure, such as $f(g(x))$ or $f(g(x) + h(y))$, going beyond examples that result from tensorization of univariate functions~\cite{oseledets_2013, shi.townsend_2021}. These functions are known to admit efficient TTN representations when the network topology reflects the underlying algebraic structure~\cite{bachmayr2021}, motivating the exploration of analogous constructions based on the simpler QTT format.

\subsection{Finite-precision QTT algebra}
\label{sec:2_qtt_framework_algebra}

\begin{figure}[t]
  \centering \includegraphics[width=\columnwidth]{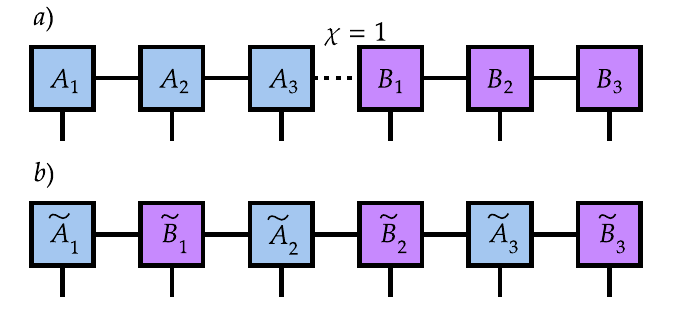}
  \caption{Illustration of the tensor product of two QTTs $\bf{A}$ and $\bf{B}$ in serial (a) and interleaved (b) orders. In the interleaved configuration (b), the cores must propagate the virtual bonds of the alternate chain, meaning the effective cores are given by $\widetilde{A}_i = A_i \otimes I$ and $\widetilde{B}_i = I \otimes B_i$, where $I$ represents the identity matrix of appropriate virtual dimension.}
  \label{fig:qtt_tensor_product}
\end{figure}

The proposed algorithm stands on an algebra of three primitive QTT operations. A fundamental primitive is the truncation of a QTT with maximum rank $r_{\bf{A}}$ to obtain another QTT of reduced size and complexity. This corresponds to approximating the optimization problem
\begin{equation}
\bf{B} = \operatorname{argmin}_{\operatorname{rank}(\bf{B}) \leq r_{\bf{B}}} \| \bf{A} - \bf{B} \|_F^2
\label{eq:qtt_truncation}
\end{equation}
where $r_{\bf{B}} < r_{\bf{A}}$ is the prescribed maximum rank for the truncated QTT $\bf{B}$. 
Several strategies exist to tackle this problem, including the TT-SVD algorithm and its variants, such as the ``zip-up'' method~\cite{stoudenmire2010, paeckel2019}. In this work, we employ an iterative truncation procedure based on a two-site variational scheme, parameterized by a tolerance $\varepsilon$ that controls the admissible residual error in the $L^2$ norm and orthogonalizes the QTT into a left- or right- orthonormal form~\cite{garcia-ripoll_2021, paeckel2019, garcia-molina.etal_2024}. This truncation primitive extends naturally to the linear combinations of QTTs by solving~\eqref{eq:qtt_truncation} for $\bf{A} = \sum_{k=1}^m \alpha_k \bf{A}_k$.

The second operation is the element-wise, or Hadamard, product of two QTTs~\eqref{eq:hadamard_product}, resulting from the multiplication of the tensor elements
\begin{equation}
    (\bf{A} \odot \bf{B})_{s_1 \ldots s_n} = \bf{A}_{s_1 \ldots s_n} \bf{B}_{s_1 \ldots s_n}.
    \label{eq:hadamard_product}
\end{equation}
This operation multiplies TT ranks of the input tensors, resulting in a maximum rank $r_{\bf{A}} r_{\bf{B}}$. To prevent rapid rank growth under repeated products, the result must be subsequently truncated. In this setting, variational truncation is advantageous over TT-SVD because its cost per sweep scales as $\O{n \ell^2 r_{\bf{A}} r_{\bf{B}} r_{\mathbf{A}\odot\mathbf{B}}^3}$, where $r_{\mathbf{A}\odot \mathbf{B}}$ is the truncated rank of the product, whereas the TT-SVD of the exact Hadamard product scales as $\O{n \ell r_{\bf{A}}^3 r_{\bf{B}}^3}$. For comparable ranks, $r_{\bf{A}} \sim r_{\bf{B}} \sim r_{\bf{A}\odot\bf{B}} \sim r$, these estimates reduce to $\O{n \ell^2 r^5}$ and $\O{n \ell r^6}$, respectively. The tradeoff is that TT-SVD is deterministic and quasi-optimal, while variational truncation is iterative and lacks the same global guarantees~\cite{paeckel2019}.

The third operation is the outer product of two QTTs, combining the vector spaces of the two tensors,
\begin{equation}
    (\bf{A} \otimes \bf{B})_{s_1 \ldots s_n t_1 \ldots t_n} := \bf{A}_{s_1 \ldots s_n} \bf{B}_{t_1 \ldots t_n}.
    \label{eq:outer_product}
\end{equation}
As shown in Fig.~\ref{fig:qtt_tensor_product}~(a), in the serial case, the QTT cores are simply concatenated, and the TT ranks remain unchanged. In contrast, for interleaved orderings, shown in Fig.~\ref{fig:qtt_tensor_product}~(b), the two QTTs must first be embedded into a common tensor space so that components associated with the same function dimension are aligned. This is achieved by inserting trivial identity tensors at the appropriate positions in the QTT representation. Subsequent Hadamard products then combine the aligned cores, resulting in a squaring of the TT ranks and requiring truncation.

\section{Orthogonal polynomial expansions in quantized tensor trains}
\label{sec:3_orthogonal}

Orthogonal polynomial bases provide a systematic framework for constructive function approximation with well-understood theoretical convergence guarantees~\cite{szeg1939}. By the Weierstrass approximation theorem, any continuous univariate function admits a convergent polynomial approximation. Moreover, as discussed in Sect.~\ref{sec:3_orthogonal-theory}, such expansions present two key properties. First, they enable an efficient and numerically stable evaluation of the polynomial approximant through three-term recurrence formulas. Second, they often present excellent convergence guarantees for smooth and highly differentiable functions, making them particularly effective.

Building on these properties, Sect.~\ref{sec:3-orthogonal-qtt} introduces a new algorithm for approximating multivariate functions in the quantized tensor-train (QTT) format. This algorithm applies orthogonal polynomial expansions to QTT arguments or operators and evaluates them efficiently using Clenshaw recurrences~\cite{clenshaw_1955}. This formulation provides a natural mechanism for function composition in QTT form, extending earlier algebraic frameworks~\cite{peng.etal_2023} and enabling efficient approximation of multivariate functions with compositional algebraic structure.

The numerical comparisons in Sect.~\ref{sec:4_results} and accompanying appendices (\ref{sec:D_comparison_1d} and~\ref{sec:E_comparison_md}) show that the method is competitive with existing QTT approximation techniques, including tensor cross-interpolation (TCI)~\cite{oseledets.tyrtyshnikov_2010,savostyanov.oseledets_2011,dolgov.savostyanov_2020,ritter.etal_2024} and multiscale interpolative constructions~\cite{lindsey_2023,chen.lindsey_2024}, avoiding direct interpolation of the full high-dimensional tensor and scaling to multivariate examples with up to 200 QTT cores.

\subsection{Orthogonal polynomial bases}
\label{sec:3_orthogonal-theory}

Any continuous univariate function $f(x)$ on a compact interval $[a, b]$ admits a convergent polynomial approximation. Orthogonal polynomial bases provide a systematic framework for constructing such approximations: orthogonality ensures that truncated expansions are uniquely defined, while the associated three-term recurrence relations enable a stable evaluation of the basis functions~\cite{szeg1939}.

Let $\{P_k\}_{k \geq 0}$ denote a family of polynomials on $[a, b]$ that are mutually orthogonal with respect to a positive weight function $\omega(x) > 0$, forming an orthogonal basis of the weighted Hilbert space $L_\omega^2([a, b])$. Truncating the expansion of $f$ in the basis after $d+1$ terms yields the degree-$d$ polynomial approximant
\begin{equation}
  f_d(x) = \sum_{k=0}^{d} c_k P_k(x),
  \label{eq:truncated_expansion}
\end{equation}
with coefficients $c_k$ uniquely determined by projection or interpolation.

This construction extends to multivariate functions via tensor-product expansions,
\begin{equation}
  f(x_1,\ldots,x_m) = \sum_{k_1=0}^{d_1}\cdots\sum_{k_m=0}^{d_m} c_{k_1\ldots k_m} \prod_{j=1}^m P_{k_j}(x_j).
  \label{eq:polynomial_multivariate}
\end{equation}
at the cost of an exponentially growing coefficient tensor $c_{k_1\ldots k_m}$. For functions with compositional algebraic structure, this cost can be avoided by representing the multivariate map as a sequence of low-dimensional polynomial expansions, each involving smaller coefficient arrays.

All classical orthogonal polynomial families satisfy three-term recurrence relations,
\begin{equation}
  P_{k+1}(x) = (\alpha_k x + \beta_k) P_k(x) - \gamma_k P_{k-1}(x), \qquad k\geq 1,
  \label{eq:polynomial_recurrence}
\end{equation}
with $\gamma_k > 0$, together with initial conditions $P_0(x)=1$ and $P_1(x)=\kappa x + \mu$ that fix the affine scaling of the polynomial sequence. This recurrence enables stable evaluation of the basis functions and, consequently, of the truncated expansion~\eqref{eq:truncated_expansion}.

Among these families, Chebyshev polynomials are particularly effective due to their strong convergence guarantees, exhibiting algebraic convergence for finitely differentiable functions and exponential convergence for analytic functions~\cite{mason.handscomb_2003,trefethen_2019}. Their expansion coefficients can be computed efficiently by interpolation at Chebyshev--Gauss or Chebyshev--Lobatto nodes, using the FFT to reduce the cost to $\O{d \log d}$, compared to the $\O{d^2}$ cost of projection-based quadrature methods. Moreover, the resulting coefficient decay provides a practical truncation criterion for guaranteeing a prescribed accuracy in the $L^\infty$ norm. In this case, the recurrence~\eqref{eq:polynomial_recurrence} specializes to $\alpha_k = 2$, $\beta_k = 0$ and $\gamma_k = 1$, with initial conditions $T_0(x)=1$ and $T_1(x)=x$. 

Taken together, these properties make Chebyshev polynomials a natural default choice for function approximation in QTT formats. Accordingly, we specialize to this basis in the numerical experiments of Sect.~\ref{sec:4_results} and in the algorithmic comparisons of~\ref{sec:D_comparison_1d} and~\ref{sec:E_comparison_md}. Our implementation is based on the SeeMPS library~\cite{garcia2026seemps}, which provides the expansion algorithm in the Chebyshev basis with support for other orthogonal bases.

\subsection{QTT composition of functions via orthogonal polynomial expansions}
\label{sec:3-orthogonal-qtt}

We now introduce an algorithm to construct the QTT encoding of a composed function $f(g(x))$, starting from a QTT encoding of the argument $g(x)$ itself. The truncated expansion in Eq.~\eqref{eq:truncated_expansion} is evaluated in QTT form by applying the basis polynomials $P_k$ element-wise to an initial QTT argument $\bf{G}$, that is, by replacing $P_k(x)$ with the QTT tensors $\bf{P}_k$ that encode the evaluation of the basis functions $P_k(g(x))$. This yields the QTT approximant 
\begin{equation}
  \bf{F}_d(\bf{G}) = \sum_{k=0}^d c_k \bf{P}_k.
  \label{eq:cal_F_qtt}
\end{equation}
A univariate non-compositional function $f(x)$ can be computed by passing the grid tensor $\bf{G} = \bf{X}$ as an argument. For a uniform grid, this tensor has an exact rank-2 QTT decomposition~\cite{khoromskij_2011}. More generally, the algorithm enables efficient construction of multivariate functions by representing the argument $g(x_1, x_2, \ldots, x_m)$ through combinations of simple QTT building blocks using outer operations described in Sect.~\ref{sec:2_qtt_framework_algebra}---such as tensor products~\eqref{eq:outer_product} or sums---with either serial or interleaved QTT core orderings.

The implementation is summarized in Algorithm~\ref{alg:clenshaw}. It assumes a fixed orthogonal polynomial basis $\{P_k\}_{k \ge 0}$ defined on an interval $[a, b]$, with known recurrence coefficients $\{\alpha_k, \beta_k, \gamma_k\}_{k\geq 1}$ and linear term $P_1(x) = \kappa x + \mu$. The expansion coefficients $c_k$ are provided as input; standard projection and interpolation procedures for computing them are described in~\cite{mason.handscomb_2003,trefethen_2019}. The algorithm also requires a rank-truncation routine $\textsc{Truncate}_\varepsilon$ parameterized by a tolerance $\varepsilon$ that controls the relative residual in the $L^2$ norm.

The QTT tensors $\bf{P}_k$ can, in principle, be constructed using the finite-precision QTT algebra of Sect.~\ref{sec:2_qtt_framework_algebra} and the three-term recurrence~\eqref{eq:polynomial_recurrence},
\begin{equation}
  \bf{P}_{k+1} = (\alpha_k \bf{G} + \beta_k \bf{1}) \odot \bf{P}_k - \gamma_k \bf{P}_{k-1},
  \label{eq:qtt_direct}
\end{equation}
involving only linear combinations and Hadamard products, where $\bf{1}$ is the QTT tensor of all ones. However, this approach is inefficient and potentially unstable, as it requires the explicit construction of all basis terms and leads to unnecessary rank growth. Instead, in practice, we compute the polynomial approximation in Eq.~\eqref{eq:cal_F_qtt} using a Clenshaw-type recursion to avoid explicit basis construction, as detailed in Algorithm~\ref{alg:clenshaw}. Numerical tests demonstrating the efficiency of this recursion compared to the direct three-term recurrence~\eqref{eq:qtt_direct} are provided in~\ref{sec:B_clenshaw}.

If the range of $g(x)$ lies in an interval $[a', b']$ different from the orthogonality interval $[a, b]$ of the basis, the argument is first rescaled to $\widetilde{\bf{G}} = \phi(\bf{G}) \in [a, b]$ using an affine bijection 
\begin{equation}
  \phi(x) = \frac{b - a}{b' - a'}(x - a') + a.
  \label{eq:affine}
\end{equation}
The provided expansion coefficients $c_k$ must correspond to the modified function $f \circ \phi^{-1}$, ensuring that the resulting approximant evaluated at $\widetilde{\bf{G}}$ represents $f(\bf{G})$ on the original domain.

\begin{figure}[t!]
  \begin{algorithm}[H]
    \caption{QTT orthogonal polynomial expansion of $f(g(x))$.}
    \label{alg:clenshaw}
    \begin{algorithmic}[1]
      \Require Expansion coefficients $\{c_k\}_{k=0}^d$ of $f \circ \phi^{-1}$, QTT representation $\bf{G}$ of $g(x)$ with known range $[a', b'] \subset \mathbb{R}$, polynomial degree $d$, TT-rank truncation tolerance $\varepsilon$, affine map $\phi : [a', b'] \mapsto [a, b]$.
      \Statex
      \Statex \textit{Affinely rescale the QTT argument using Eq.~\eqref{eq:affine}}
      \State $\widetilde{\bf{G}} \gets \phi(\bf{G})$ \mbox{  \quad \qquad $\phi: [a', b'] \to [a, b]$}
      \Statex
      \Statex \textit{Evaluate Clenshaw recursion using Eq.~\eqref{eq:qtt_clenshaw}}
      \State $n \gets $ \mbox{number of QTT cores in $\bf{G}$}
      \State $\bf{1} \gets$ \mbox{QTT tensor of all ones with $n$ cores}
      \State $\bf{Y}_{d+2}, \ \bf{Y}_{d+1} \gets$ \mbox{zero QTT with $n$ cores}
      \For{$k=d$ \textbf{to} 1}
        \State $\bf{Y}_k \gets
          c_k\bf{1} + \bf{R}_k \odot \bf{Y}_{k+1}
          - \gamma_{k+1}\bf{Y}_{k+2}$
        \State $\bf{Y}_k \gets \textsc{Truncate}_\varepsilon(\bf{Y}_k)$

      \EndFor
      \State $\bf{P}_1 \gets \kappa \widetilde{\bf{G}} + \mu \bf{1}$
      \State \Return $\textsc{Truncate}_\varepsilon \left(c_0 \bf{1} + \bf{P}_1 \odot \bf{Y}_1 - \gamma_1 \bf{Y}_2\right)$
    \end{algorithmic}
  \end{algorithm}
\end{figure}

The computational cost is dominated by the truncations following Hadamard products. As discussed in Sec.~\ref{sec:2_qtt_framework_algebra}, the two-site variational truncation of $\bf{A} \odot \bf{B}$ to an output QTT with rank $r_{\bf{A}\odot \bf{B}}$ has a typical per-sweep cost $\O{n\ell^2 r_{\bf A}r_{\bf B}r_{\bf{A}\odot\bf{B}}^3}$. Hence, for comparable input and output ranks, this gives a cost $\O{n \ell^2 r^5}$ per sweep, instead of the $\O{n \ell r^6}$ cost of TT-SVD truncation of the exact Hadamard product. 

For the direct recurrence in Eq.~\eqref{eq:qtt_direct}, the variational truncation of each step $(\alpha_k\bf G+\beta_k\bf 1) \odot \bf{P}_k$ has typical per-sweep cost $\O{n\ell^2 r_{\bf G} r_{\bf P_k} r_{\bf G\odot\bf P_k}^3}$. Using the pessimistic rank bound $r_{\bf P_k}\leq k+1$ for univariate polynomials with $\ell=2$, and assuming comparable intermediate ranks, gives the naive worst-case cost estimate $\sum_{k=0}^{d} \O{n r_{\bf G} (k+1)^4} = \O{n r_{\bf G}d^5}$. The Clenshaw recursion in Algorithm~\ref{alg:clenshaw} has the same leading structure, with $\bf{P}_k$ replaced by the intermediate tensors $\bf Y_{k+1}$. 

In practice, however, these bounds are overly conservative: truncation strongly reduces the intermediate ranks, and these are highly nonuniform across QTT cores. This explains the substantially better empirical scaling observed in the benchmarks. In multivariate settings, however, no analogous degree-only rank bound is available, and intermediate tensors may accumulate large TT ranks despite repeated variational truncation.

From a numerical standpoint, enforcing an orthonormal QTT form during truncation significantly improves robustness compared to non-orthonormal schemes. In our experiments, the dominant source of numerical instability arises from extreme function values in high-dimensional settings, primarily due to the limitations of finite-precision floating-point arithmetic.

\section{Numerical benchmarks}
\label{sec:4_results}

In this section, we evaluate the performance of the QTT polynomial expansion algorithm on a collection of univariate and multivariate functions, focusing on the Chebyshev basis. As figures of merit, we report the error in discretized $L^{\infty}$ norm, corresponding to the maximum pointwise error; the maximum QTT rank $r_\text{max}$, serving as a proxy for the memory footprint $\Theta(2 r_\text{max}^2)$; and the wall-clock time in seconds. These quantities are reported as functions of the number $n$ of QTT cores, the polynomial degree $d$, and the truncation tolerance $\varepsilon$. We use TT-SVD to construct quasi-optimal QTT representations from exact vector arrays, and use the variational truncation procedure to compress intermediate QTT Hadamard products within the Chebyshev expansion. Both steps are controlled by the tolerance $\varepsilon$. A systematic assessment requires multiple independent experiments, which we organize in three-column figures with shared horizontal axes and standardized vertical ranges.

We have additionally compared the performance of the algorithm with tensor cross-interpolation (TCI)~\cite{oseledets.tyrtyshnikov_2010} and multiscale interpolative constructions~\cite{lindsey_2023}. For clarity, univariate and multivariate results are reported separately in~\ref{sec:D_comparison_1d} and~\ref{sec:E_comparison_md}, respectively.

\subsection{Univariate functions}
\label{sec:4_results_univariate}

\begin{figure}[t]
  \centering%
  \includegraphics[width=0.9\columnwidth]{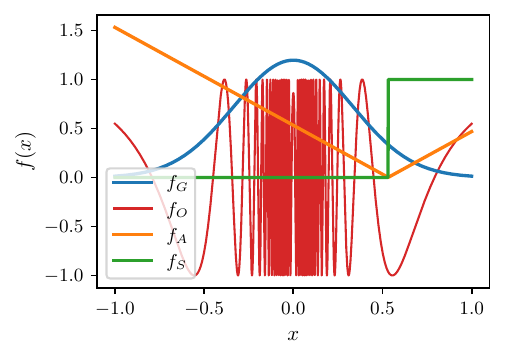}
  \caption{Functions considered for the univariate benchmarks: Gaussian distribution $f_G$~\eqref{eq:3_func_gaussian} (blue), oscillatory function $f_O$~\eqref{eq:3_func_osc} (red), absolute value function $f_A$~\eqref{eq:3_func_abs} (orange) and Heaviside step function $f_S$~\eqref{eq:3_func_step} (green). The color code is used consistently throughout the benchmarks.}
  \label{fig:functions}
\end{figure}

We benchmark the QTT expansion algorithm on a set of representative univariate functions, shown in Fig.~\ref{fig:functions}. First, we consider a Gaussian distribution
\begin{equation}
  f_G(x) = \frac{1}{\sigma \sqrt{2 \pi}} e^{-\left(\frac{x}{2\sigma}\right)^2}
  \label{eq:3_func_gaussian}
\end{equation}
with $\sigma = 1/3$, and an oscillatory function
\begin{equation}
  f_O(x) = \cos \left(\frac{1}{x^2 + \varepsilon}\right)
  \label{eq:3_func_osc}
\end{equation}
with $\varepsilon=10^{-2}$. Both functions are analytic and are therefore expected to exhibit exponential convergence with the polynomial order $d$, following an initial error plateau.

To assess the performance on non-smooth targets, we additionally consider the absolute value function
\begin{equation}
  f_A(x) = |x - x_c|,
  \label{eq:3_func_abs}
\end{equation}
and the Heaviside step function
\begin{equation}
  f_S(x) = \Theta(x - x_c),
  \label{eq:3_func_step}
\end{equation}
centered at $x_c = \frac{1}{2} + 2^{-5}$ to avoid artificially low-rank representations arising from symmetry. Since the Heaviside step function is discontinuous, its Chebyshev expansion exhibits persistent Gibbs oscillations and does not converge uniformly. Similarly, the absolute value function possesses only first-order derivatives of bounded variation, resulting in slow algebraic convergence.

\begin{figure*}[t]
  \centering
  \includegraphics[width=\textwidth]{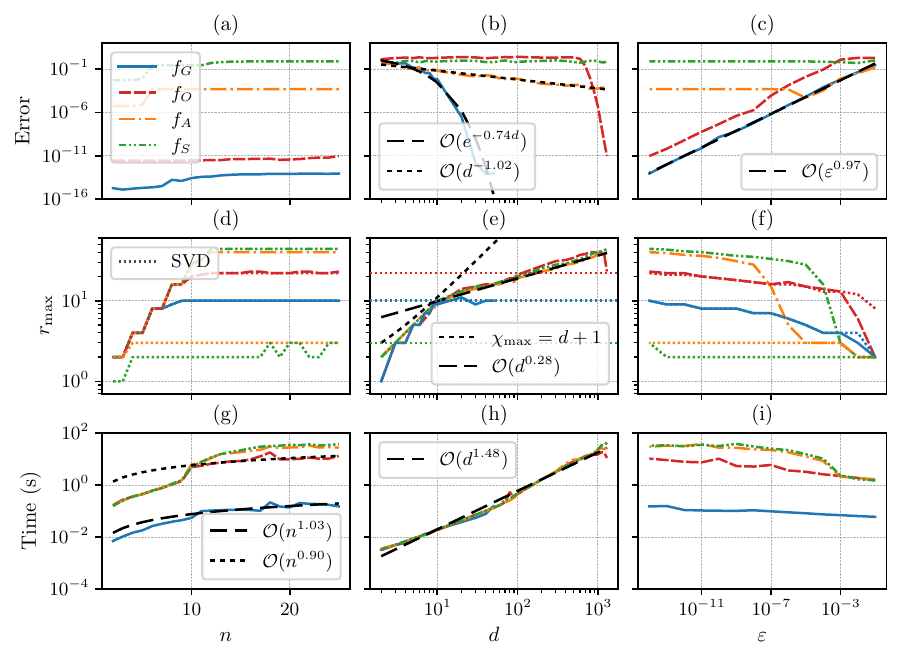}
  \caption{Benchmark results for the QTT Chebyshev expansion of the univariate functions in Fig.~\ref{fig:functions}. Each column fixes two parameters and varies the third, with shared vertical and horizontal axes.
  Left: varying $n$ with fixed $d=1300$, $\varepsilon=10^{-14}$. 
  Center: varying $d$ with fixed $n=25$, $\varepsilon=10^{-14}$. 
  Right: varying $\varepsilon$ with fixed $n=25$, $d=1300$. 
  For the Gaussian function, the polynomial order is set to $d_G=50$ to optimize performance.}
  \label{fig:chebyshev_1d}
\end{figure*}

\begin{figure*}[t]
  \centering%
  \includegraphics[width=\textwidth]{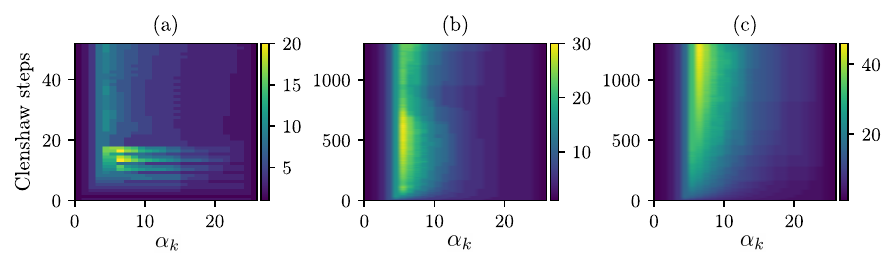}
  \caption{Heat map of TT-rank evolution during Clenshaw evaluations with $n=25$ cores and $\varepsilon = 10^{-14}$. The horizontal axis corresponds to the contracted indices $\alpha_k$ [cf.~Eq.~\eqref{eq:tt}], while the vertical axis indexes the $d$ Clenshaw recursion steps. The color scale indicates the TT-rank size, illustrating intermediate tensor sizes and controlled rank growth. Panel (a) shows the Gaussian function $f_G(x)$ with $d=50$, while panels (b) and (c) show the oscillatory $f_O(x)$ and absolute value $f_A(x)$ functions with $d=1300$.}
  \label{fig:chebyshev_1d_ranks}
\end{figure*}

The benchmarks explore up to $25$ QTT cores, corresponding to grids with $2^{25}\approx 10^{7}$ points. The truncation tolerance is varied down to $\varepsilon=10^{-14}$, yielding errors close to machine precision. The polynomial order $d$ is increased up to $d=1300$ except for the Gaussian function, which requires only $d_G=50$; as shown in~\ref{sec:B_clenshaw}, overestimating the polynomial order can degrade performance.

Figure~\ref{fig:chebyshev_1d} summarizes the results, reporting the approximation error (top row), maximum TT rank (middle row), and wall-clock time (bottom row) as functions of the number of QTT cores (first column), polynomial order (second column), and truncation tolerance (third column). 

The approximation error remains largely independent of the grid resolution, even for the highly oscillatory function, since increasing the number of grid points only samples the same fixed polynomial approximant more densely, with the dominant error set by the polynomial degree $d$. The maximal TT ranks saturate to function-dependent values, reflecting the limited complexity of polynomials (see Figs.~\ref{fig:chebyshev_1d}(a, d)). These ranks closely match the quasi-optimal ranks obtained via direct TT-SVD. The computational cost (see Fig.~\ref{fig:chebyshev_1d}(g)) scales linearly with the number of cores, with an empirical rate of $\mathcal{O}(n^{0.97\pm 0.07})$, indicating that a high density of interpolation is achievable with very few additional resources, in line with results provided by alternative expansion schemes~\cite{lindsey_2023}. This is further confirmed by the rank decay at every intermediate Clenshaw step in Fig.~\ref{fig:chebyshev_1d_ranks}, a plot that captures the size of the tensors in the QTT expansion as a function of the evaluation steps.

Fig.~\ref{fig:chebyshev_1d}(b) reflects a convergence rate that follows the theoretical expectations. Moreover, Fig.~\ref{fig:chebyshev_1d}(e) shows that the maximum TT rank scales algebraically with the polynomial order as $\O{d^{0.28}}$ for all test functions. This curve lies systematically below the pessimistic upper bound of $\chi_\text{max} = d+1$ known for $d$-th order polynomials (see~\ref{sec:A_polynomials}). This exponent should not be interpreted as a universal rank-scaling law; rather, it is an empirical finite-range fit that reflects the compressibility of the sampled polynomial approximants under the prescribed TT truncation tolerance. At convergence, this rank is consistent with the quasi-optimal TT-SVD value, as shown for the Gaussian and oscillating functions for $d_G=50$ and $d=1300$, respectively. Finally, Fig.~\ref{fig:chebyshev_1d}(h) shows how the wall-clock time scales at an algebraic rate of $\O{d^{1.48}}$ with the polynomial order, well below the naive worst case scenario of $\O{d^4}$. Again, this is attributed to the intermediate rank profiles of the QTT approximations shown in Fig.~\ref{fig:chebyshev_1d_ranks}. As the QTT cores are highly rectangular, the expected $\O{d^3}$ cost for TT-SVD is subleading.

Finally, we have confirmed that the TT-rank truncation tolerance parameterizes both the approximation error, TT ranks, and wall-clock time, as expected (see Figs.~\ref{fig:chebyshev_1d}(c, f, i)). In particular, the error is a monotonically decreasing function of $\varepsilon$ up until it is dominated by the expansion error. Furthermore, the maximal TT ranks increase mildly---logarithmically with the tolerance for all functions, enabling a tunable tradeoff between accuracy and computational cost when a good polynomial approximation is warranted.

\begin{figure*}[t]
  \centering \includegraphics[width=0.75\textwidth]{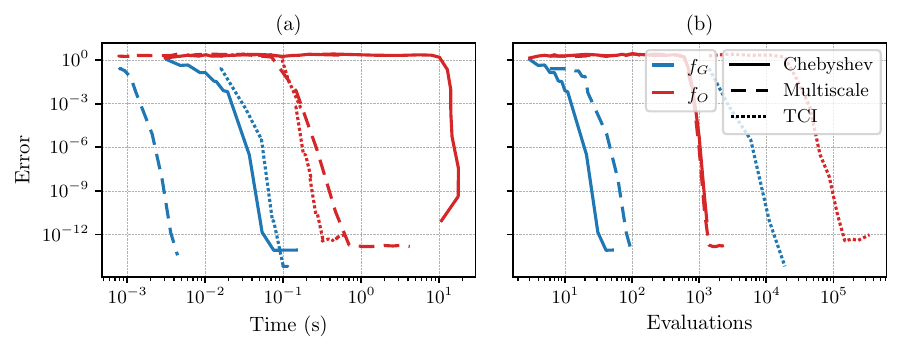}
  \caption{Performance comparison for loading the Gaussian $f_G(x)$~\eqref{eq:3_func_gaussian} and oscillating $f_O(x)$~\eqref{eq:3_func_osc} functions in QTT format, using QTT Chebyshev expansions (solid), rank-revealing interpolative constructions (dashed), and TCI (dotted). (a) Error versus wall-clock time. (b) Error versus number of function evaluations.}
  \label{fig:benchmark_1d}
\end{figure*}

Orthogonal polynomial expansions remain competitive with other QTT approximation techniques for smooth univariate functions. Figure~\ref{fig:benchmark_1d} compares QTT Chebyshev expansions with TCI and multiscale interpolative constructions, with benchmark curves respectively extracted from the detailed comparisons reported in~\ref{sec:D_comparison_1d_tci} and~\ref{sec:D_comparison_1d_multiscale}. 

We restrict this comparison to the Gaussian~\eqref{eq:3_func_gaussian} and oscillatory~\eqref{eq:3_func_osc} functions, since polynomial methods are not expected to be competitive with TCI for non-smooth targets. In particular, TCI is not constrained by the convergence limitations of global polynomial approximation and therefore provides a substantial advantage for discontinuous or non-smooth functions. The results are reported in Fig.~\ref{fig:benchmark_1d}. Figures~\ref{fig:benchmark_1d}(a) and~(b) report the $L^\infty$ error against wall-clock time and number of function evaluations, respectively. All three methods converge exponentially for the smooth targets considered, but with different cost profiles. Multiscale interpolative constructions generally converge faster than QTT Chebyshev expansions while requiring a comparable number of function evaluations, determined by the polynomial order $d$. TCI is especially effective for the oscillatory function, due to its faster convergence with the rank cap $r_{\mathrm{cap}}$, despite requiring substantially more function evaluations.

Overall, QTT Chebyshev expansions exhibit comparable performance for smooth univariate functions while requiring far fewer function evaluations than TCI, by approximately two orders of magnitude in these benchmarks. These results confirm that orthogonal polynomial expansions provide an effective and structure-preserving framework for loading highly differentiable univariate functions. The logarithmic scaling with discretization size $n$, due to the TT-rank decay at intermediate expansion steps, demonstrates scalability for highly dense approximations. Moreover, its linear scaling with polynomial order $d$ demonstrates robust scalability to large orders, aligning with theoretical error guarantees for smooth and highly differentiable functions. This motivates its extension to multivariate problems through function-composition approaches.

\subsection{Multivariate functions}
\label{sec:4_results_multivariate}

\begin{figure*}[t]
  \centering
  \begin{minipage}{\textwidth}
    \includegraphics[width=0.75\linewidth]{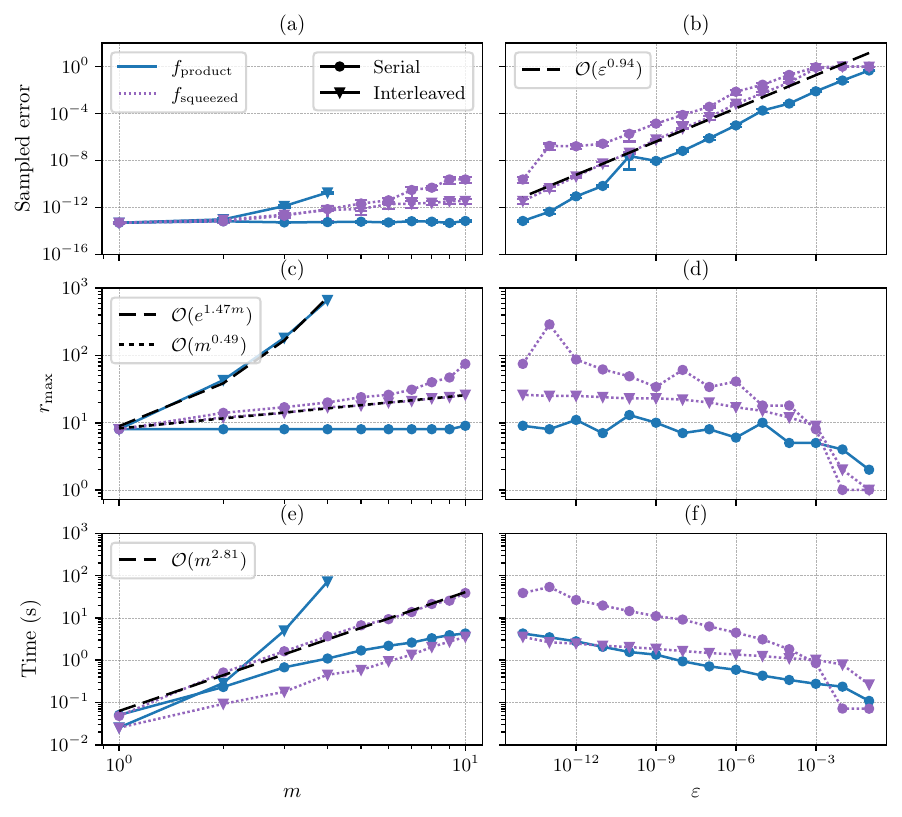}
    \caption{QTT Chebyshev expansions of the product Gaussian $\G_{\mathrm{product}}(\mathbf{x})$~\eqref{eq:4_gaussian_product}~(blue) and squeezed Gaussian $\G_{\mathrm{squeezed}}(\mathbf{x})$~\eqref{eq:4_gaussian_squeezed}~(purple). Circular and triangular markers, respectively, denote serial and interleaved QTT core orderings. The left column fixes the truncation tolerance at $\varepsilon=10^{-14}$, varying dimension. The right column fixes the function dimension at $m=10$, varying the tolerance. All scenarios use $n=20$ QTT cores per dimension and an optimally-chosen polynomial order $d$.}
    \label{fig:chebyshev_md}
  \end{minipage}
\end{figure*}

\begin{figure*}[t]
  \centering
  \begin{minipage}{\textwidth}
    \includegraphics[width=0.75\linewidth]{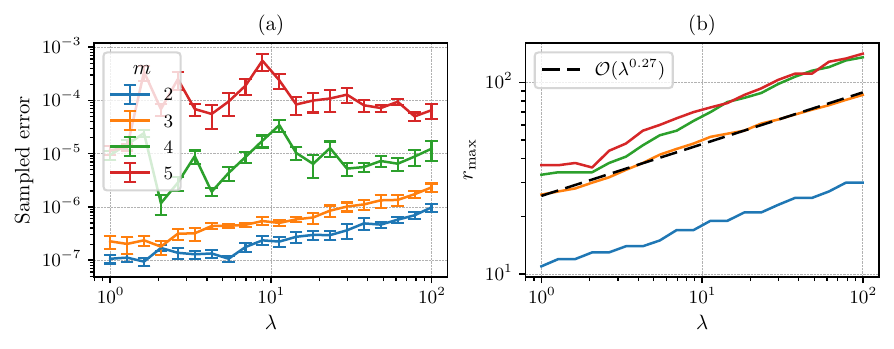}
    \caption{QTT expansions of the quartic model~\eqref{eq:quartic_potential}. Panel (a) shows the sampled error as a function of the coupling strength $\lambda \in [1, 100]$ and dimension $m=2,\ldots,5$. Panel (b) reports the corresponding maximum TT ranks. All results use $n=14$ QTT cores per dimension in serial ordering and are obtained with the truncation tolerance $\varepsilon=10^{-8}$.}
    \label{fig:quartic}
  \end{minipage}
\end{figure*}

We now demonstrate the ability of orthogonal polynomial expansions to encode multivariate functions. In general, such expansions follow from the tensor-product representation in Eq.~\eqref{eq:polynomial_multivariate}. A direct strategy computes the full coefficient tensor and expands it into QTT form; however, the exponential growth in coefficients---and the associated increase of QTT operations and Clenshaw evaluations---limits this approach to low-dimensional problems. Alternatively, one may approximate the function using TCI, but this method is non-deterministic, lacks explicit convergence guarantees, and may require a large number of function evaluations (see~\ref{sec:E_comparison_md}). Here, we advocate a compositional strategy: when the function presents compositional structure, it can be encoded in QTT form by chaining low-dimensional polynomial expansions, avoiding constructing the full coefficient tensor, and enabling accuracy control through the TT-rank truncation tolerance~$\varepsilon$, with a drastic reduction of function evaluations compared to TCI.

We first consider two multivariate Gaussian-like functions, whose QTT compressibility is guaranteed by previous studies in the literature~\cite{rohrbach2020}. The first is an unnormalized product of Gaussians
\begin{equation}
  \G_\text{product}(\mathbf{x}) = \exp \left( -\sum_{i=1}^m x_i^2 \right).
  \label{eq:4_gaussian_product}
\end{equation}
The second example is a ``squeezed'' Gaussian-like function
\begin{equation}
  \G_\text{squeezed}(\mathbf{x}) = \exp\left( -\left[\sum_{i=1}^m x_i\right]^2 \right),
  \label{eq:4_gaussian_squeezed}
\end{equation}
which, despite its apparent strong correlations, actually admits a compact QTT representation. Both functions result from the composition of an univariate function $f(x) = \exp(-x)$ with simple multivariate building blocks, namely $\sum_i x_i^2$ and $\left(\sum_i x_i\right)^2$, which can be efficiently assembled in QTT form using outer operations, described in Section~\ref{sec:2_qtt_framework_algebra}.

We study these functions for dimensions $m = 1, \ldots, 10$, varying the truncation tolerance $\varepsilon$ between $10^{-1}$ and $10^{-14}$ to explore accuracy-performance tradeoffs. The largest configurations involve up to 200 QTT cores and approximately $10^{60}$ tensor elements, rendering direct evaluation of the discretized pointwise error infeasible. Instead, the error is estimated via random sampling. Given $S$ grid points $\{\mathbf{x}^{(j)}\}_{j=1}^S$ drawn uniformly from the tensor-product grid, the sampled error of a QTT approximation $\mathbf{F}_d$ to $f(\mathbf{x})$ is defined as
\begin{equation}
  \max_{1 \leq j \leq S} \left| f(\mathbf{x}^{(j)}) - \mathbf{F}_d(\mathbf{x}^{(j)}) \right|,
  \label{eq:sampled_error}
\end{equation}
where $\mathbf{F}_d(\mathbf{x}^{(j)})$ denotes the evaluation of the QTT tensor at the sampled grid point. In this work, we report the average of Eq.~\eqref{eq:sampled_error} over 10 independent trials with $S=1000$ samples each. As discussed in Sect.~\ref{sec:C_sampling}, this procedure captures the overall error trends reliably.

The results are summarized in Fig.~\ref{fig:chebyshev_md}. These plots show the approximation error (top row), maximum tensor rank (middle row), and CPU wall-clock time (bottom row), as a function of the number of variables $m$ and the truncation tolerance $\varepsilon$, comparing serial and interleaved QTT core orderings.

The multidimensional polynomial expansion exhibits excellent convergence properties, as shown in Fig.~\ref{fig:chebyshev_md}(a) for $m=1$ to 10 variables and all tensor orders. The approximation errors are, as before, dominated by the truncation tolerance $\varepsilon$, until they reach the theoretical error of the polynomial expansion.
As shown in Figs.~\ref{fig:chebyshev_md}(c, e), compressibility and performance are highly sensitive to the tensor order. The product Gaussian~\eqref{eq:4_gaussian_product} exhibits rapidly growing complexity under interleaved ordering, while the squeezed Gaussian~\eqref{eq:4_gaussian_squeezed} remains efficiently representable in both orderings, achieving algebraic TT-rank scaling of approximately $\mathcal{O}(m^{0.49})$. The sampled approximation errors in Fig.~\ref{fig:chebyshev_md}(a) correlate with the observed ranks, while the wall-clock time in Fig.~\ref{fig:chebyshev_md}(e) scales polynomially with dimension. Notably, the squeezed Gaussian displays improved stability and gentler rank growth under interleaved ordering.

The right column of Fig.~\ref{fig:chebyshev_md} illustrates the effect of the truncation tolerance $\varepsilon$ for the most stringent setting ($n=20$, $m=10$). The product Gaussian is omitted in the interleaved ordering due to its prohibitive complexity. As shown in Figs.~\ref{fig:chebyshev_md}(b, d), the tolerance parameter controls accuracy and complexity in a manner analogous to the univariate benchmarks of Fig.~\ref{fig:chebyshev_1d}, enabling reductions in wall-clock time by orders of magnitude at the cost of increased truncation error (Fig.~\ref{fig:chebyshev_md}(f)).

We next consider a nonlinear model that lies beyond the structured Gaussian-like regime and admits no a priori rank estimates. Specifically, we study the quartic model
\begin{equation}
  \G_{\mathrm{quartic}}(\mathbf{x}) = \exp \left(- \left[\sum_{i=1}^m x_i^4 + \lambda \sum_{j < i} x_i^2 x_j^2 \right] \right),
  \label{eq:quartic_potential}
\end{equation}
where the coupling parameter $\lambda \geq 0$ controls the strength of dense pairwise interactions. For $\lambda=0$, the model reduces to a product of independent univariate factors, while increasing $\lambda$ induces progressively stronger correlations. Despite its nonlinear structure, this model remains compatible with the QTT polynomial framework: the multivariate polynomial inside the exponential is assembled from low-rank univariate components using additions and Hadamard products, after which the nonlinearity is handled via a univariate polynomial expansion of $f(x)=\exp(-x)$. 

The experiment fixes $n=14$ QTT cores per dimension on the symmetric domain $x_i \in [-1,1]$, with truncation tolerance fixed to $\varepsilon=10^{-8}$ and adaptively chosen polynomial order. The largest configurations involve up to 70 QTT cores, corresponding to more than $10^{21}$ tensor elements. We restrict to the serial QTT core ordering, which consistently outperforms the interleaved alternative for this model. The results are summarized in Fig.~\ref{fig:quartic}.

The sampled error shown in Fig.~\ref{fig:quartic}(a) increases mildly with the coupling strength $\lambda$ and monotonically with the dimension $m$, reflecting the growing nonlinearity and correlation of the model. These errors exceed the prescribed tolerance $\varepsilon$, since the variational truncation bounds the approximation error in the $L^2$ norm, whereas the reported errors correspond to the $L^\infty$ (pointwise) norm and can therefore be significantly larger. The maximal TT ranks in Fig.~\ref{fig:quartic}(b) grow algebraically with both parameters, leading to sub-exponential increases in computational complexity and enabling the treatment of higher-dimensional, strongly coupled systems.

Overall, these experiments demonstrate that QTT orthogonal polynomial expansions efficiently approximate multivariate functions with intrinsic low rank through function composition, achieving subexponential complexity and overcoming the curse of dimensionality. Moreover, they highlight the critical role of QTT core ordering and show that, even for nonlinear and densely coupled models, the proposed framework maintains scalability in moderate dimensions.

\section{Discussion}
\label{sec:5_discussion}
This work has explored the problem of computing quantized tensor train representations of univariate and multivariate functions, motivated by their applications in \emph{quantum-inspired} numerical analysis. To this end, an algorithm based on the expansion of univariate functions on orthogonal polynomial bases has been developed. The approach relies on a finite-precision algebra of QTT operations, including linear combinations and Hadamard products, which enables the translation of classical polynomial expansions and Clenshaw evaluation schemes into the QTT framework.

Without loss of generality, the framework has been exhaustively tested on the Chebyshev basis. The numerical results show that, despite the finite-precision arithmetic induced by the repeated rank truncations required by the QTT algebra, the polynomial expansions are stable, achieve the expected theoretical convergence rates, and scale using polynomial computational resources. In particular, in the univariate case, the constructions scale linearly with both the polynomial order and the number of QTT cores---i.e., with a logarithmic dependence on the grid density. This favorable scaling extends to multivariate functions, which can likewise be encoded using polynomial resources in low-rank scenarios.

Beyond function approximation, QTT polynomial expansions provide a generic framework for function composition, enabling nonlinear manipulation of initial QTT arguments. These capabilities may be useful for the solution of nonlinear equations within quantum-inspired methods~\cite{lubasch.etal_2018}, such as for the implementation of filters in Hamiltonian problems~\cite{yang.etal_2020}.

A natural generalization of the method consists of adapting direct tensorization schemes (see Eq.~\eqref{eq:polynomial_multivariate}) to QTT formats. This would enable expanding multivariate functions defined on several QTT arguments, increasing the range of compositional structures that can be approximated. However, the corresponding tensor of coefficients and number of required rank truncations scales exponentially with the total number of variables, restricting this approach to low-dimensional settings.

As shown by the comparisons reported in~\ref{sec:D_comparison_1d} and~\ref{sec:E_comparison_md}, current implementations of the QTT expansion framework do not systematically outperform state-of-the-art alternatives for univariate function loading. Nevertheless, the method exhibits favorable asymptotic scaling, which can be advantageous in regimes where alternative approaches become less efficient. In addition, the performance-accuracy tradeoff enabled by the truncation tolerance provides an effective mechanism for computing accurate approximations with controlled computational cost.

The algorithms and methods presented in this work are freely available in the SeeMPS library~\cite{garcia2026seemps}, an open-source Python framework designed to facilitate the prototyping and experimentation of QTT and quantum-inspired algorithms. All simulations reported here were performed on a workstation equipped with an 18-core Intel(R) Xeon(R) W-2295 CPU and 64GB of RAM. The data generated in this study, together with the source files required to reproduce the numerical experiments, are available from the authors upon reasonable request.

\section*{Acknowledgments}
  This work has been supported by Spanish Projects No. PID2021-127968NB-I00 and No. PDC2022-133486-I00, funded by MCIN/AEI/10.13039/501100011033 and by the European Union “NextGenerationEU”/PRTR. The authors acknowledge the discussions on related topics with A. Bou Comas and E. Staffetti. The authors also gratefully acknowledge the Scientific Computing Area (AIC), SGAI-CSIC, for their assistance while using the DRAGO Supercomputer for performing the simulations.  The authors thank the Institut Henri Poincaré (UAR 839 CNRS-Sorbonne Université) and the LabEx CARMIN (ANR-10-LABX-59-01) for their support. PGM acknowledges the funds given by ``FSE invierte en tu futuro" through an FPU Grant FPU19/03590 and by MCIN/AEI/10.13039/501100011033.

\appendix
\section{Exact QTT polynomial encodings}
\label{sec:A_polynomials}
\begin{figure}[t]
  \centering
  \includegraphics[width=0.85\columnwidth]{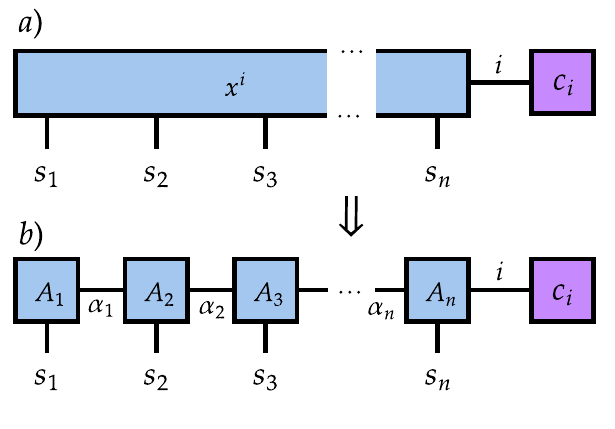}
  \caption{Decomposition of a parameterized collection of monomials $x^i$ contracted with a vector of coefficients $c_i$ in a QTT.}
  \label{fig:A_polynomial}
\end{figure}

This appendix provides a recipe to construct the QTT representation of degree $d$ polynomials $f_d(x) = \sum_{i=0}^d c_i x^i$ over a finite domain $x\in[a,b]$ on a grid of arbitrary density. The constructive proof creates QTT cores $A_{i,r}^s \in \mathbb{C}^{\ell\times (d+1)^2}$ whose ranks are determined by the polynomial degree $r=d+1$ and the quantization $\ell$.

The proof proceeds by induction. Starting from $\bf{X}^i_n$, the QTT decomposition of a monomial $x^i$ using $n$ cores, the construction derives a recipe for $\bf{X}^i_{n+1}$ by adding one more core,
\begin{equation}
  \bf{X}^i_{n+1} = \sum_{j=0}^d A_{j, i}^{s_{n+1}}\bf{X}^j_n.
\end{equation}
The new discretization can be expressed in terms of the original, coarser-grained one $x(s_1,s_2,\ldots, s_n)$ and a displacement that depends on the new quantization variable $s_{n+1}\in\{0,1,\ldots, \ell-1\}$
\begin{equation}
  x(\mathbf{s}_{n+1}) = x(\mathbf{s}_n) + \frac{d}{\ell^{n+1}}s_{n+1}.
\end{equation}
Hence, it follows
\begin{align}
  x(\mathbf{s}_{n+1})^i
    & = \sum_{j=0}^i x(\mathbf{s}_n)^j \left(\frac{s_{n+1}}{\ell^{n+1}}\right)^{i-j} {{i}\choose{j}}            \\
    & =\sum_{j=0}^d x(\mathbf{s}_n)^j \left(\frac{s_{n+1}}{\ell^{n+1}}\right)^{i-j} {{i}\choose{j}} \Theta(i-j) \\
    & =:\sum_{j=0}^d x(\mathbf{s}_n)^j A_{j,i}^{s_{n+1}}.\label{eq:recursion}
\end{align}
This expression uses the asymmetric Heaviside function, which is nonzero $\Theta(i-j)=1$ only when $i\geq j$. Equation~\eqref{eq:recursion} provides a recursive formula to construct the tensor representation of $\bf{X}^i_n$
\begin{equation}
  \bf{X}^i_n = \sum_{\alpha_1, \ldots, \alpha_{n-1}} A_{0,\alpha_1}^{s_1}A_{\alpha_1,\alpha_2}^{s_2}\cdots A_{\alpha_{n-1},i}^{s_n}
  \label{eq:monomials-A}
\end{equation}
with tensors
\begin{equation}
  A_{j,k}^{s_n} = \left(\frac{s_n}{\ell^n}\right)^{k-j}{{k}\choose{j}}\Theta(k-j),
\end{equation}
where the indices $j,k$ and $\alpha_i$ run between 0 and $d$. 

As illustrated in Fig.~\ref{fig:A_polynomial}(a), this construction results in a QTT with one free index of size $d + 1$. These constructions extend to arbitrary polynomials of a fixed degree by contracting their coefficients $\{c_i\}_{i=0}^d$ with that free index. Thus, for a degree-$d$ polynomial, we obtain a QTT decomposition with TT ranks of size $d+1$
\begin{equation}
  f_d(x) = \sum_{i=0}^d c_i x^i\;\Rightarrow\; \bf{F}_{d,n} = \sum_{i=0}^d c_i \bf{X}^i_n,
\end{equation}
following Eq.~\eqref{eq:monomials-A}, as shown in Fig.~\ref{fig:A_polynomial}(b).

\section{Clenshaw evaluation of QTT polynomial expansions}
\label{sec:B_clenshaw}
\begin{figure*}[t]
  \centering
  \includegraphics[width=0.85\textwidth]{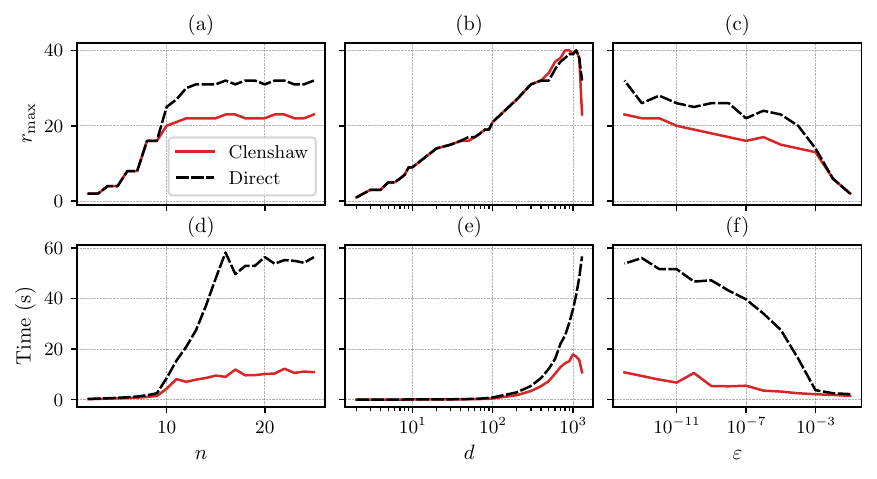}
  \caption{Performance comparison for the evaluation of QTT Chebyshev expansions~\eqref{eq:qtt_direct} using Clenshaw~\eqref{eq:qtt_clenshaw} (in red) and direct (in black) evaluation formulas for the oscillating function $f_O(x)$~\eqref{eq:3_func_osc}.}
  \label{fig:B_clenshaw_1}
\end{figure*}
\begin{figure}[t]
  \centering
  \includegraphics[width=0.8\columnwidth]{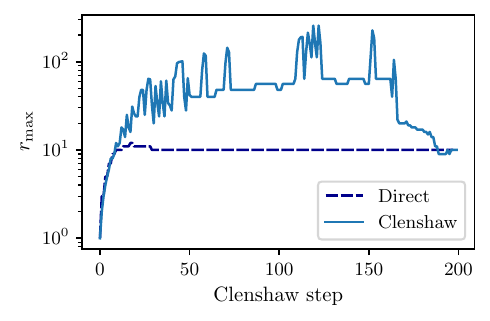}
  \caption{Maximum TT ranks reached during the $d$ intermediate evaluations of the QTT Chebyshev expansion of the Gaussian~\eqref{eq:3_func_gaussian} with $n=25$, $d=200$ and $\varepsilon=10^{-14}$, considering both direct and Clenshaw evaluation steps.}
  \label{fig:B_clenshaw_2}
\end{figure}

The Clenshaw evaluation method exploits the three-term recurrence relation of the polynomial basis~\eqref{eq:qtt_direct} to evaluate the expansion without explicitly forming each basis polynomial, reducing intermediate rank growth and improving numerical stability and performance~\cite{clenshaw_1955}. To this end, it introduces auxiliary QTT tensors initialized as $\bf{Y}_{d+2}=\bf{Y}_{d+1}=0$, and computed backwards according to the recursion
\begin{align}
  \bf{Y}_{d+2} &= \bf{Y}_{d+1} = 0 \\
  \bf{Y}_{k} &= c_{k} \bf{1} + (\alpha_k \bf{X} + \beta_k \bf{1}) \odot \bf{Y}_{k+1} - \gamma_{k+1} \bf{Y}_{k+2},
  \label{eq:qtt_clenshaw}
\end{align}
with each step followed by rank truncation to control rank growth. The polynomial approximant is then recovered as
\begin{equation}
  \bf{F}_d = c_0 \bf{1} + \bf{P}_1 \odot \bf{Y}_1 - \gamma_1 \bf{Y}_2,
\end{equation}
where $\bf{1}$ is the all-ones QTT and $\bf{P}_1 = \kappa \bf{X} + \mu \bf{1}$. This avoids the explicit reconstruction of all basis tensors $\bf{P}_k$, limiting the propagation of high-rank intermediate representations.

Figure~\ref{fig:B_clenshaw_1} compares the performance of direct and Clenshaw-based evaluations for the oscillatory function~\eqref{eq:3_func_osc}, using the most stringent parameters $n=25$, $d=1300$ and $\varepsilon = 10^{-14}$. Both approaches achieve similar accuracy (not shown for brevity), but Clenshaw evaluations consistently yield lower wall-clock times and significantly smaller TT ranks.

Interestingly, overestimating the polynomial order $d$ degrades the performance and stability of Clenshaw evaluations. Figure~\ref{fig:B_clenshaw_2} shows this behavior for the Gaussian function, where the polynomial order is set to a large overestimate of $d=200$, and QTT Clenshaw evaluations show higher instability and larger intermediate ranks. However, since the optimal order in the Chebyshev basis can be estimated reliably from the decay of the expansion coefficients, this behavior does not pose a practical problem.

\section{QTT Monte Carlo error estimation}
\label{sec:C_sampling}

\begin{figure*}[t]
  \centering
  \includegraphics[width=0.8\textwidth]{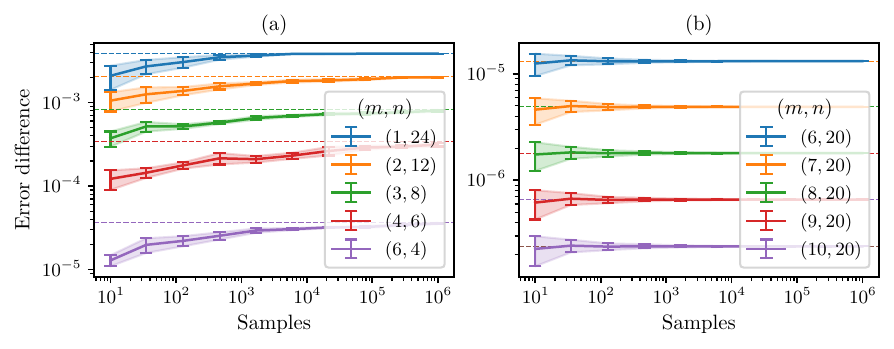}
  \caption{Convergence of sampled error estimates for a product Gaussian. (a) Sampled $L^{\infty}$ error compared with the exact full-tensor error. (b) Sampled $L^2$ error compared with the exact contraction-based QTT error. The legend reports the dimension $m$ and QTT cores per dimension $n$, including the standard deviation over 100 repetitions.}
  \label{fig:D_sampling}
\end{figure*}

Direct evaluation of $L^p$ distances between multivariate functions is generally infeasible at high dimensions due to the cost of accessing the full discretized tensor. We therefore assess the reliability of Monte Carlo error estimation in two representative settings.

First, we test whether the sampled $L^\infty$ error converges to the exact truncation error induced by a TT-SVD approximation of a product Gaussian. Fig.~\ref{fig:D_sampling}(a) shows convergence of the sampled maximum error as the number of samples increases, across several dimensions and QTT resolutions.

Second, we consider larger QTTs for which the full tensor error is inaccessible and instead use the exact contraction-based $L^2$ error as a reference. As shown in Fig.~\ref{fig:D_sampling}(b), the sampled $L^2$ error converges reliably to this reference with a moderate number of samples.

Although the exact asymptotic convergence of the sampled errors to the true error is not assessed, the sampled estimates are consistently observed to be of the correct order of magnitude across all tested configurations. Since the sampled errors are used here primarily as a practical diagnostic to compare algorithms and assess relative performance, rather than to certify tight error bounds, this level of accuracy is sufficient for our purposes. 

\section{Univariate benchmarks}
\label{sec:D_comparison_1d}

This appendix reproduces the univariate QTT expansion benchmark developed in Section~\ref{sec:4_results_univariate} for two representative state-of-the-art algorithms: multiscale interpolative constructions and tensor cross-interpolation (TCI). The same test functions, parameter sweeps, and performance metrics are employed to enable a direct and consistent comparison.

\subsection{Multiscale interpolative constructions}
\label{sec:D_comparison_1d_multiscale}

\begin{figure*}[t]
  \centering
  \includegraphics[width=\textwidth]{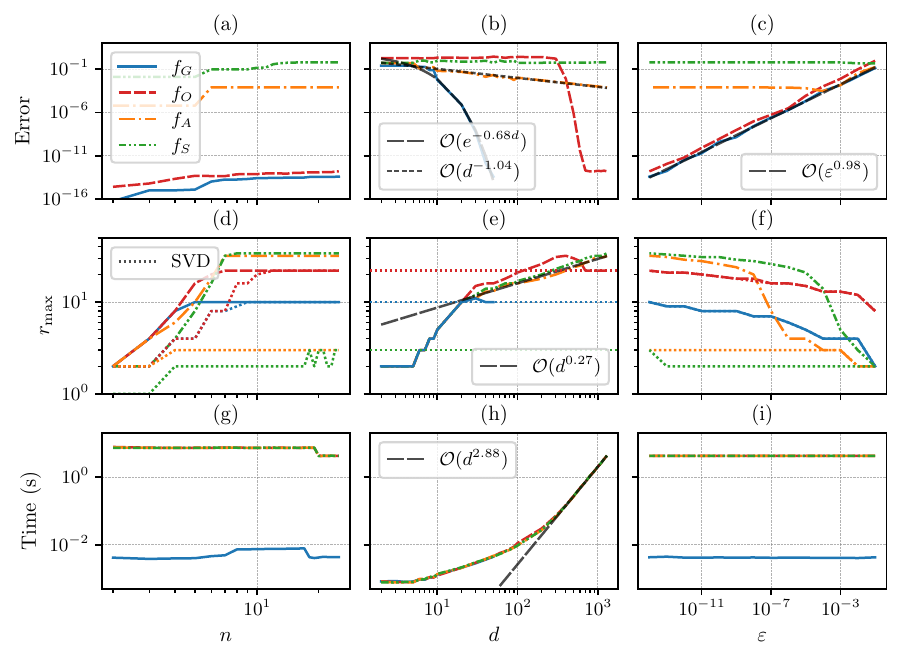}
  \caption{Multiscale interpolations of the four functions shown in Fig.~\ref{fig:functions}, following the same structure as Fig.~\ref{fig:chebyshev_1d}. Each column fixes two parameters and varies a third.
  Left: vary $n$ with fixed $d=1300$, $\varepsilon=10^{-14}$. 
  Center: vary $d$ with fixed $n=25$, $\varepsilon=10^{-14}$. 
  Right: vary $\varepsilon$ with fixed $n=25$, $d=1300$. 
  The polynomial order for the Gaussian function is set to $d_G = 50$ to enhance convergence.}
  \label{fig:3_lagrange_1d}
\end{figure*}

Multiscale interpolative constructions, as developed by Lindsey~\cite{lindsey_2023,chen.lindsey_2024}, provide a structured approach for building QTT representations of polynomial interpolants based on Chebyshev--Lagrange interpolation. The method constructs the QTT representation in a single forward pass by assembling tensor cores that encode interpolation on two nested Chebyshev--Lobatto grids, each using $d+1$ points. A distinctive feature of this approach is that only the leftmost QTT core depends on the target function, while all remaining cores are universal and can be reused across different interpolants.

In its rank-revealing variant, the construction applies local SVD truncation during assembly to control the QTT ranks, avoiding a costly global compression step. Fig.~\ref{fig:3_lagrange_1d} reports a benchmark analogous to that of Fig.~\ref{fig:chebyshev_1d}, varying the grid resolution $n$, interpolation order $d$, and SVD truncation tolerance $\varepsilon$. Figs.~\ref{fig:3_lagrange_1d}(a–c) show that multiscale interpolative constructions achieve convergence rates comparable to those of QTT Chebyshev expansions, consistent with the theoretical expectations.

By construction, the computational cost is nearly independent of the number of QTT cores $n$, as shown in Fig.~\ref{fig:3_lagrange_1d}(g), since all bulk cores except the leftmost one are equivalent. In contrast, QTT Chebyshev expansions exhibit linear growth with $n$. However, as shown in Fig.~\ref{fig:3_lagrange_1d}(h), multiscale interpolative constructions display unfavorable scaling with the interpolation order, with an accelerating trend and a fitted power-law complexity of $\mathcal{O}(d^{2.88})$, whereas QTT Chebyshev expansions scale nearly linearly with $d$. Finally, Fig.~\ref{fig:3_lagrange_1d}(i) shows that the method is largely insensitive to the SVD tolerance $\varepsilon$, as truncation affects only local Schmidt coefficients rather than the global construction cost.

A fundamental limitation of this approach is its lack of scalability to multivariate settings. Since the method relies on explicit function sampling, the associated tensor of samples grows exponentially with dimension, similarly to the coefficient tensor in direct multivariate polynomial constructions [cf.~Eq.~\eqref{eq:polynomial_multivariate}]. This exponential scaling renders the method impractical beyond low-dimensional problems.

\subsection{Tensor cross-interpolation}
\label{sec:D_comparison_1d_tci}

\begin{figure*}[t]
  \centering
  \includegraphics[width=\textwidth]{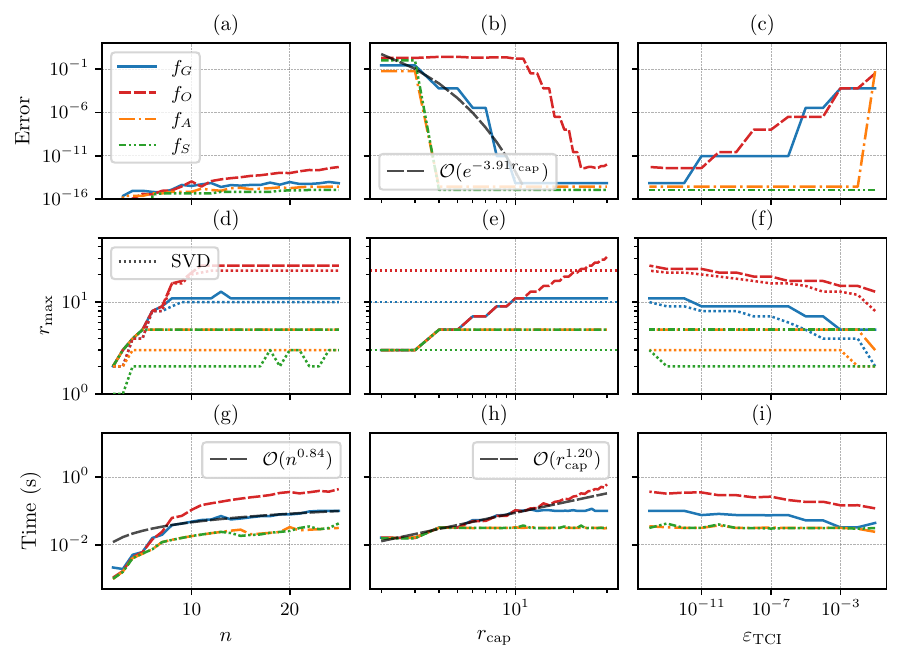}
  \caption{Tensor cross-interpolation of the four functions shown in Fig.~\ref{fig:functions}, following the same structure as Fig.~\ref{fig:chebyshev_1d}. Each column fixes two parameters and varies a third.
  Left: vary $n$ with fixed $r_{\mathrm{cap}}=30$, $\varepsilon=10^{-14}$.
  Center: vary $r_{\mathrm{cap}}$ with fixed $n=25$, $\varepsilon=10^{-14}$.
  Right: vary $\varepsilon$ with fixed $n=25$, $r_{\mathrm{cap}}=30$.}
  \label{fig:3_cross_1d}
\end{figure*}

Tensor cross-interpolation (TCI) is a well-known technique for encoding black-box functions in QTT format by sampling its elements~\cite{oseledets.tyrtyshnikov_2010,savostyanov.oseledets_2011,dolgov.savostyanov_2020,ritter.etal_2024}. Unlike previous algorithms, TCI does not build a polynomial or algebraic structure; instead, it samples the target function (or tensor) to obtain the QTT cores that minimize a specified cost function. Essentially, it iteratively refines an initial QTT representation of the target function following ``sweeps'' that increase its ranks until convergence is achieved. In our implementation, based on~\cite{chertkov.etal_2022}, the TT ranks grow by 2 with each sweep and are capped by a predefined threshold $r_{\mathrm{cap}}$, which is interpreted as the analogue of the polynomial order $d$. Notably, as TCI does not rely on a polynomial approximation, it is not constrained by the smoothness of the target function and can efficiently approximate non-smooth functions.

This study replicates the benchmark of Fig.~\ref{fig:chebyshev_1d} up to a TT rank $r_{\mathrm{cap}}=30$ and a tolerance $\varepsilon_{\text{TCI}}$ for which the algorithm halts. The results are presented in Fig.~\ref{fig:3_cross_1d}. As shown in Figs.~\ref{fig:3_cross_1d}(a) and~\ref{fig:3_cross_1d}(b), TCI converges to machine precision for all test functions, including non-differentiable targets, exhibiting exponential convergence and significantly outperforming the previous polynomial methods, with a fitted rate of $\mathcal{O}(e^{-3.91r_{\mathrm{cap}}})$ for the Gaussian target. Figs.~\ref{fig:3_cross_1d}(d) and~\ref{fig:3_cross_1d}(e) show how it slightly overestimates the TT ranks relative to those computed with TT-SVD. Moreover, Figs.~\ref{fig:3_cross_1d}(g) and~\ref{fig:3_cross_1d}(h) illustrate that the sublinear asymptotic time scaling of TCI, of order $\mathcal{O}(n^{0.84})$, is comparable to the linear rate of QTT expansions (see Fig.~\ref{fig:chebyshev_1d}) and the constant rate of multiscale interpolative constructions (see Fig.~\ref{fig:3_lagrange_1d}).

\FloatBarrier

\section{Multivariate benchmarks}
\label{sec:E_comparison_md}

\begin{figure*}[t]
  \centering
  \includegraphics[width=0.8\textwidth]{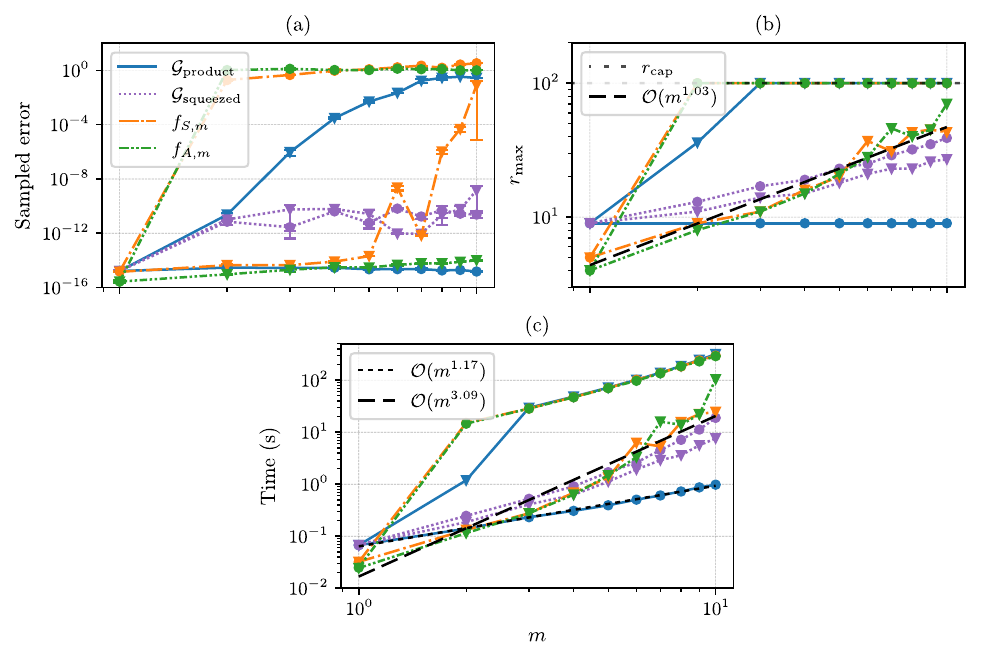}
  \caption{TCI approximations of the Gaussian-like functions $\mathcal{G}_{\mathrm{product}}$~\eqref{eq:4_gaussian_product}~(blue) and $\mathcal{G}_{\mathrm{squeezed}}$~\eqref{eq:4_gaussian_squeezed}~(purple), including the multivariate non-smooth functions $f_{A,m}(\mathbf{x})$~\eqref{eq:4_abs_multivariate}~(orange) and $f_{S,m}(\mathbf{x})$~\eqref{eq:4_step_multivariate}~(green). The function dimension $m$ varies from 1 to 10, with $n=20$ QTT cores per dimension. This follows a similar layout as Fig.~\ref{fig:chebyshev_md}, with comparable axis ranges to facilitate direct comparison.}
  \label{fig:cross_md}
\end{figure*}

\begin{figure*}[t]
  \centering
  \includegraphics[width=0.8\textwidth]{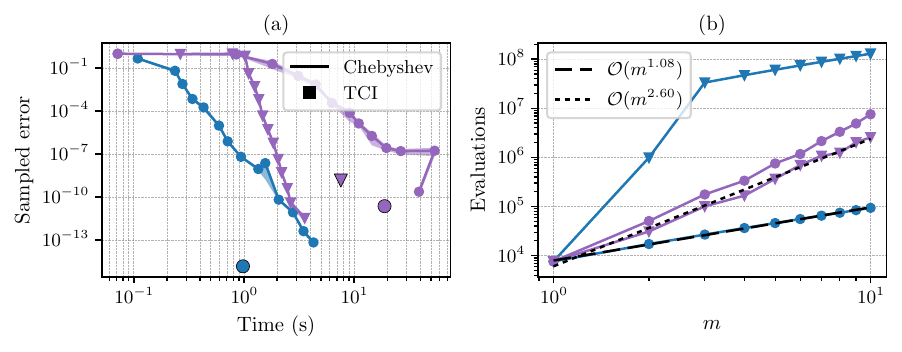}
  \caption{Performance comparison between QTT Chebyshev expansions (see Fig.~\ref{fig:chebyshev_md}) and TCI (see Fig.~\ref{fig:cross_md}). Panel~(a) reports wall-clock time versus sampled $L^\infty$ error for Gaussian-like functions in both serial and interleaved QTT orders. Chebyshev results appear as curves parameterized by $\varepsilon$, while TCI results are shown as isolated markers. Markers and colors consistently identify the three test functions, enabling a direct comparison of the performance of TCI and polynomial expansions across all cases. Panel~(b) reports the number of function evaluations required by TCI to converge, depending on the function dimension. The results for the product Gaussian in interleaved order are artificially capped by $r_{\mathrm{cap}}$.}
  \label{fig:benchmark_md}
\end{figure*}

This appendix reproduces the multivariate QTT expansion benchmarks developed in Section~\ref{sec:4_results_multivariate} using TCI. We consider an increasing dimension $m$, between 1 and 10, with $n=20$ QTT cores per dimension, and estimate the error through random sampling. As TCI increases the ranks in a sweeping fashion, the maximum TT ranks are fine-tuned to ensure an accurate convergence.

As an aside, since TCI is able to encode non-smooth functions efficiently, we include two $m$-dimensional extensions of the absolute-value $f_A$~\eqref{eq:3_func_abs} and step $f_S$~\eqref{eq:3_func_step} functions, given by
\begin{align}
  f_{A, m}(\mathbf{x}) & = \left|\sum_{i=1}^m x_i - x_c\right|,\mbox{ and}
  \label{eq:4_abs_multivariate}                                                 \\
  f_{S, m}(\mathbf{x}) & = \Theta\left(\sum_{i=1}^m x_i - x_c\right).
  \label{eq:4_step_multivariate}
\end{align}
shown respectively in green and orange colors, with threshold $x_c = \tfrac{1}{2} + 2^{-5}$.

The results are included in Fig.~\ref{fig:cross_md}, confirming that TCI can efficiently load the Gaussian-like functions when the QTT core ordering is chosen appropriately. In these cases, the TT ranks scale sublinearly with the dimension. Consequently, the required computational resources scale polynomially: linearly for the product Gaussian function and as $\mathcal{O}(m^{2.81})$ for the squeezed functions in both core orders, as shown in Fig.~\ref{fig:cross_md}(c). Moreover, TCI can accurately load the non-smooth functions $f_{A, m}$~\eqref{eq:4_abs_multivariate} and $f_{S, m}$~\eqref{eq:4_step_multivariate}. Interestingly, for these functions, the interleaved order yields an exponentially better compression, enabling near-optimal convergence with only polynomial computational resources.

Since QTT polynomial expansions are parameterized by the truncation tolerance $\varepsilon$, we can compare their performance against TCI by varying this accuracy-cost tradeoff. Fig.~\ref{fig:benchmark_md}(a) compares the wall-clock time required to reach a given sampled error by both approaches. Since TCI is not parameterized, its results appear as single-shot evaluations, whereas the QTT results trace a curve parameterized by $\varepsilon$. These results show how QTT expansions can achieve lower wall-clock times than TCI for all functions, with acceptable accuracy. Moreover, for the squeezed Gaussian in interleaved order, QTT expansions clearly outperform TCI in terms of accuracy and performance. TCI also requires several orders of magnitude more function evaluations than QTT expansions: as illustrated in Fig.~\ref{fig:benchmark_md}(b), TCI requires up to $10^7$ evaluations, as opposed to the $\O{d}$ evaluations required by polynomial expansions. This large difference may become performance-critical when evaluations are costly.

\clearpage

\bibliographystyle{apsrev4-2}
\bibliography{bibliography}

\end{document}